\documentclass
[showpacs,superscriptaddress,prl,twocolumn,tightenlines,balancelastpage,10pt,a4paper]{revtex4}%

\usepackage{graphicx}
\usepackage{subfigure}
\usepackage{dcolumn}
\usepackage{bm}

\begin{document}
\preprint{To be revised}
\title{Demonstration of efficient scheme for generation of ``Event Ready" entangled photon pairs from single photon source
}

\author{Qiang Zhang}
\affiliation{Hefei National Laboratory for Physical Sciences at
Microscale \& Department of Modern Physics, University of Science
and Technology of China, Hefei, Anhui 230026, P.R. China}
 \affiliation{Physikalisches Institut,
Universit\"{a}t Heidelberg, Philosophenweg 12, 69120 Heidelberg,
Germany}
\author{Xiao-Hui Bao}
\affiliation{Hefei National Laboratory for Physical Sciences at
Microscale \& Department of Modern Physics, University of Science
and Technology of China, Hefei, Anhui 230026, P.R. China}
\author{Chao-Yang Lu}
\affiliation{Hefei National Laboratory for Physical Sciences at
Microscale \& Department of Modern Physics, University of Science
and Technology of China, Hefei, Anhui 230026, P.R. China}
\author{Xiao-Qi Zhou}
\affiliation{Hefei National Laboratory for Physical Sciences at
Microscale \& Department of Modern Physics, University of Science
and Technology of China, Hefei, Anhui 230026, P.R. China}
\author{Tao Yang}
\affiliation{Hefei National Laboratory for Physical Sciences at
Microscale \& Department of Modern Physics, University of Science
and Technology of China, Hefei, Anhui 230026, P.R. China}
\author{Terry Rudolph}
\affiliation{Optics Section, Blackett Laboratory, Imperial College
London, London SW7 2BZ, United Kingdom}
\affiliation{Institute for
Mathematical Sciences, Imperial College London, London SW7 2BW,
United Kingdom}
\author{Jian-Wei Pan}
\affiliation{Hefei National Laboratory for Physical Sciences at
Microscale \& Department of Modern Physics, University of Science
and Technology of China, Hefei, Anhui 230026, P.R. China}
 \affiliation{Physikalisches Institut,
Universit\"{a}t Heidelberg, Philosophenweg 12, 69120 Heidelberg,
Germany}

\begin{abstract}
We present a feasible and efficient scheme, and its
proof-of-principle demonstration, of creating entangled photon
pairs in an event-ready way using only simple linear optical
elements and single photons. The quality of entangled photon pair
produced in our experiment is confirmed by a strict violation of
Bell's inequality. This scheme and the associated experimental
techniques present an important step toward linear optics quantum computation.

\end{abstract}
\pacs{03.67.Mn, 03.65.Ud, 03.67.-a} \maketitle

There is a considerable worldwide effort to produce clean single
photon sources; whole conferences \cite{conference} and journal
issues \cite{njp} have been devoted to the topic. Leading
technologies in this effort are based on physical systems as
diverse as quantum dots \cite{qdot}, nitrogen vacancy centers in
diamond \cite{nvcentre}, single trapped atoms \cite{atom},
filtered signal photons from parametric down-conversion
\cite{fransonetc} and surface acoustic waves in silicon
\cite{saw}, to name but a few. Although they will find immediate
uses in quantum communication, one of the more exciting
possibilities for single photon sources is that they may be
utilized with only linear optical elements and photon number
detectors to build a quantum computer, as was shown in the seminal
paper of Knill, Laflamme and Milburn \cite{KLM}.

A first necessary step to turn single photons into a useful
resource for quantum information processing is to generate ``event
ready" entangled pairs from them. Although the KLM scheme shows
that this is in principle possible, in practise their proposal
requires keeping complicated interferometers stable over a photon
wavelength; moreover the gates succeed with probability only 1/16.
By contrast we present here a new idea for generation of entangled
photon pairs from single photons which requires stability only
over the coherence length of the photons, and which succeeds with
a probability of up to 3/16 \cite{rudolph}. We demonstrate this
idea using filtered signal photons from down conversion
\cite{fransonetc}.

\begin{figure}
\includegraphics[width=2.5in]{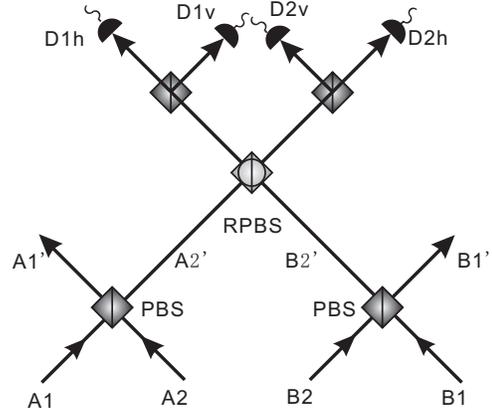}
\caption{\label{fig:figure1} Scheme for creation of ``event ready"
entangled photon pairs from single photons. RPBS represents a
45-degree oriented PBS.}
\end{figure}

Our scheme for generation of a maximally entangled pair of photons
given four single photons as input is shown in Fig.
\ref{fig:figure1}, which works as follows:

Four single photons A1,A2,B1,B2 are each prepared in the state $|H\rangle+|V\rangle$, corresponding to polarization of 45-degrees. Photons A1 and A2 interfere at one polarizing beam
splitter (PBS), B1 and B2 interfere at
another PBS. The state after the two PBS's is
\begin{eqnarray}
|\psi\rangle=&&
\frac{1}{4}(|HV\rangle_{A1'}|0\rangle_{A2'}+|0\rangle_{A1'}|HV\rangle_{A2'}\nonumber\\
&&+|H\rangle_{A1'}|H\rangle_{A2'}+
 |V\rangle_{A1'}|V\rangle_{A2'})
\nonumber\\
&&\otimes(|HV\rangle_{B1'}|0\rangle_{B2'}+|0\rangle_{B1'}|HV\rangle_{B2'}\nonumber\\&&+
|H\rangle_{B1'}|H\rangle_{B2'}+|V\rangle_{B1'}|V\rangle_{B2'})
\end{eqnarray}

The outputs A2' and B2' then undergo Type-II fusion \cite{rudolph} - specifically they are interacted at a PBS oriented at 45-degrees (accomplished by inserting one half-wave plate in
each of the two inputs and two outputs of an ordinary PBS), and then undergo polarization sensitive detection. Whenever there is a coincidence between detectors D1 (either D1h or
D1v) and D2 (either D2h or D2v), photon A1' and B1' will be
collapsed into a maximally entangled Bell state. To understand why, note from Eq.~1 that the four-photon state has 16 terms
before entering the 45-degree oriented PBS. However, the photons
in the state of $|HV\rangle|0\rangle$ or $|0\rangle|HV\rangle$
will `` stick" together because of the ``photon bunching" effect
\cite{grangier} when passing through the 45-degree oriented PBS.
 Therefore, choosing the conditions where there is one
 and only one photon in each photon-number detector D1 and D2,
 i.e. a coincidence between D1 and D2, we have post-selected the
 items $(|H\rangle_{A1'}|H\rangle_{A2'}+
 |V\rangle_{A1'}|V\rangle_{A2'})(|H\rangle_{B1'}|H\rangle_{B2'}+|V\rangle_{B1'}|V\rangle_{B2'})$
 from Eq.~1.

An alternative way to see what happens is as follows. Similar to the case of ``entanglement swapping" \cite{swapping, jianwei},
we can rewrite the state after the initial two PBS's as:
\begin{eqnarray}
|\psi\rangle&=&|\phi^{+}\rangle_{A1'A2'}\otimes|\phi^{+}\rangle_{B1'B2'}\nonumber\\
&=&|\psi^{+}\rangle_{A2'B2'}\otimes|\psi^{+}\rangle_{A1'B1'}\nonumber\\
&+&|\psi^{-}\rangle_{A2'B2'}\otimes|\psi^{-}\rangle_{A1'B1'}\nonumber\\
&+&|\phi^{+}\rangle_{A2'B2'}\otimes|\phi^{+}\rangle_{A1'B1'}\nonumber\\
&+&|\phi^{-}\rangle_{A2'B2'}\otimes|\phi^{-}\rangle_{A1'B1'}.
\end{eqnarray}

To generate an entangled photon pair in A1' and B1' mode, two PBSs
are placed after the 45-degree oriented PBS to make a partial
Bell-state measurement. When there is a coincidence between D1h
and D2h, or between D1v and D2v, the state of photon A2' and
photon B2' must be $|\phi^{+}\rangle_{A2'B2'}$ and the state of
photon A1', B1' will be collapsed into
$|\phi^{+}\rangle_{A1'B1'}$. Alternatively, coincidence between D1h and
D2v or between D1v and D2h demonstrates that photon A2', B2' are
in the state of $|\psi^{+}\rangle_{A2'B2'}$ , which will collapse
the photon A1', B1' in a state of $|\psi^{+}\rangle_{A1'B1'}$
\cite{probability}.

\begin{figure}
\includegraphics[width=3.5in]{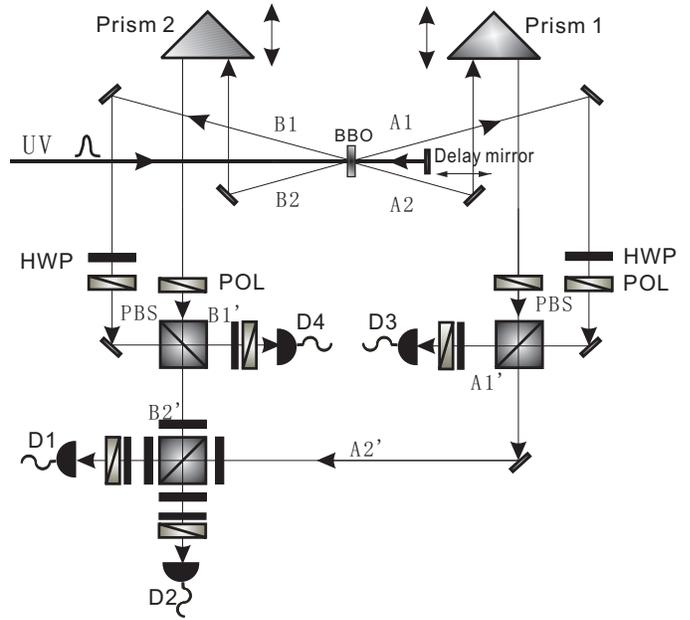}
\caption{\label{fig:figure2} Experimental setup for a
proof-of-principle demonstration of our scheme. A UV pulse with a
length of 180fs passes through a $2$ mm BBO($\beta-barium
borate$) crystal and produces a polarized entangled photon pair
via the process of type II SPDC. The UV pulse is reflected by a
delay mirror mounted on a step motor(minimum step is $0.1 \mu$m)
to pass the BBO crystal twice and prepare the second entangled
photon pairs. Half wave plates(HWP) and polarizers(Pol.) are used
to disentangle the photon pairs and to generate single photons.
Prism 1 and 2 are all mounted on micrometer translation
stages (minimum step is $10\mu$ m) to adjust the path length. Four
45-degree HWPs and a normal PBS constitute the 45-degree oriented
PBS. Two 0 degree polarizers are placed before Detector 1 and 2 to
make a partial Bell-state measurement instead of PBSs suggested in
our original scheme. The photons are collected by fiber couplers
into avalanched diode single photon detectors after narrow band
filters (with a FWHM of 2.8 nm).}
\end{figure}

In our experiment, the four single photons are achieved by
filtering signal photons from type II spontaneous parametric
down-conversion(SPDC) \cite{kwiat}. The setup of our experiment is
shown in Fig.~\ref{fig:figure2}. A 394nm ultra violet(UV) pulse
passes through a nolinear crystal (BBO) twice and generates two
polarization entangled photon pairs in the state
$|\phi^{+}\rangle=\frac{1}{\sqrt{2}}(|H\rangle|H\rangle+|V\rangle|V\rangle$,
via  SPDC. The power of the UV pulse is 600 mw and
produces 10,000 pairs per second. Four 45-degree
linear polarizers are utilized to disentangle the photon pairs
into single photons(A1-2, B1-2) each in the state
 $|+\rangle=\frac{1}{\sqrt{2}}(|H\rangle+|V\rangle)$. To make
itself a perfect single photon source, photon A1(B1) needs to take
their twin photon A2(B2) as a trigger signal and vice versa
\cite{fransonetc}, which means photon A1(B1) cannot be seen as a
single photon source until photon A2(B2) is detected. Therefore,
to guarantee that the photon sources are real single photon
source, four photon coincidences are necessary in our
experiment. The four photon coincidence also has another
advantage: In Eq. 2, only the desired item
$|\phi^{+}\rangle_{A2'B2'}|\psi^{+}\rangle_{A2'B2'}$ or
$|\psi^{+}\rangle_{A2'B2'}|\psi^{+}\rangle_{A1'B1'}$  can provide
a four-fold coincidence between detectors D1(D1h or D1v), D2(D2h
or D2v), D3 and D4, so the four-fold coincidence can help to post
select the desired state instead of photon number detection in the
original scheme.

With the help of a prism 1(2) mounted on a micrometer translation
stage, single photon A1(B1) and A2(B2) are interfered at a PBS as
suggested in Fig.~1. Scanning the prism 1(2)'s position to
overlap the input single photons perfectly at the PBS's, we can
achieve an "Hong-Ou-Mandle" type interference  curve\cite{ou}. We
lay the prism 1(2) on the position which provides the best
interference visibility (about 94\%).

\begin{figure}
\includegraphics[width=3.5in]{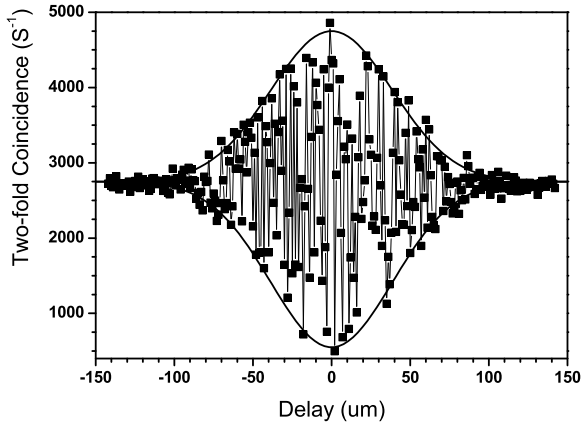}
\caption{\label{fig:figure3} Interference fringe observed when the
delay mirror is moved to achieve perfect temporal overlap. We
measure the two-fold coincidence between the output modes toward
detectors D1 and D2 behind 45 degree polarizers, by scanning the
position of the delay mirror with step sizes of 1 $\mu m$. The
envelope of the observed twofold coincidence varies indicating the
visibility of the two-photon coherence. Inside the coherent
region, the best visibility is obtained at the position where
perfect temporal overlap is achieved.}
\end{figure}

After the interference, the output modes A2', B2' are directed into the
45-degree oriented PBS as is shown in Fig. \ref{fig:figure2}. To
further generate the ``event ready" entangled photon pairs, we
vary B2's path length by scanning the Delay Mirror such that
photon A2' and B2' arrive at the 45-degree oriented PBS
simultaneously. However, due to the poor four-fold coincidence
(about 0.3 per minute) and the ultrashort coherence length of the
single photons (about $200 \mu m$), it is not easy to achieve a
good spatial and temporal overlap of  photons A2' and B2' at the
45-degree oriented PBS. As such, we developed an easier way to achieve the
interference with a slight modification of the setup. We change the
two half-wave plates at the input modes of the 45-degree oriented
PBS from 45 degrees to 0 degrees. With the modification, the ``photon
bunching'' effect will disappear and the orthogonally polarized
photons in the terms $|HV\rangle_{A2'}$ or $|HV\rangle_{B2'}$ will
be separated by the 45-degree oriented PBS, and will give a coincidence
between Detectors 1 and 2. When the two modes A2' and B2' are
overlapped perfectly and a coincidence between D1 and D2 is
observed, the states of the two output modes will be collapsed
into $(|H\rangle|V\rangle +
e^{i\phi}|V\rangle|H\rangle)/\sqrt{2}$, where $\phi$ is the phase
difference of the two input path modes. As is shown in Fig.
\ref{fig:figure3}, we scan the delay mirror with a step motor to
observe the two-photon interference curve and lock the delay at
the position with the best visibility, which is just the position
for the photons to perfectly overlap. Since only two photon are
involved in the above process, the coincidence is much higher than
than the four-photon case and it is also much easier to find the
interference position.

Obviously, to give a proof-of-principle demonstration of the
working principle of our scheme, it is sufficient  only to identify
the state $|\phi^{+}\rangle_{A2'B2'}|\phi^{+}\rangle_{A1'B1'}$.
Therefore, in the experiment, we put two 0-degree polarizers
respectively in  front of D1 and D2, instead of the PBSs in the
idealized scheme, which are utilized to discriminate
$|\phi^{+}\rangle_{A2'B2'}$ and $|\psi^{+}\rangle_{A2'B2'}$. Only the state $|\phi^{+}\rangle_{A2'B2'}$ can
provide a coincidence between D1 and D2.

\begin{figure}
\includegraphics[width=3in]{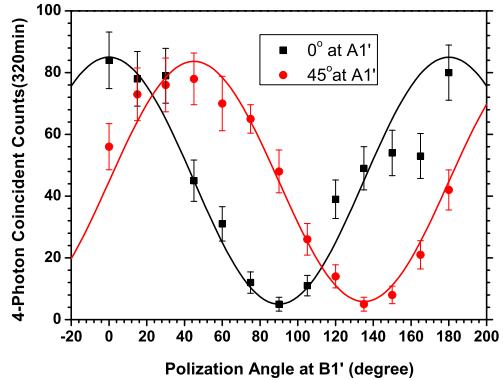}
\caption{ \label{fig:figure4} Entanglement verification.
Coincidence between photon A1' and B1' conditioned on the
detection of photon A2' and photon B2'. When setting photon B1' at
0 degree or 45 degree respectively and varying the polarizer angle
A1', the two sine curves with a visibility of $89\%$  demonstrate
that the two single photon A1' and B1' are really entangled.}
\end{figure}

Upon projection of photon A2' and photon B2' into the
$|V\rangle|V\rangle$ state, photon A1' and B1' should be in the
state $|\phi^{+}\rangle_{A1'B1'}$. To verify whether this
entangled state is obtained or not, we analyze the polarization
correlation between photon A1' and photon B1' conditioned on
coincidences of D1 and D2. We utilize two polarizers in modes A1'
and B1' to analyze the polarization coherence. B1's polarizer is
put at 0 or 45 degree, and we change the polarizer in A1's mode to
do the analysis. If the entangled photon pair is produced, the
twofold coincidence between A1' and B1' should show two sine
curves as functions of $\theta_{A1'}$, as $\theta_{B1'}$ is set at
0 or 45 degree respectively. Fig. \ref{fig:figure4} shows the
 experimental result for the coincidences between A1' and B1', given that photons A2' and B2' have been
 registered as a trigger. The experimentally obtained coincidences shown in Fig. \ref{fig:figure4}
 have been fitted by a joint sine function with the same amplitude for both curves. The observed
 visibility of $89\%$ clearly surpasses the 0.71 limit of Bell's inequality \cite{Bell}, which indicates
  the photon pair is genuinely entangled.

The high-visibility sinusoidal coincidence curves in the experiment
imply a violation of the Clauser-Horne-Shimony-Holt
(CHSH)\cite{chsh} inequality, $S<2$ for any local
theory, where $S=E(a,b)-E(a,b^{'})+E(a^{'},b)+E(a^{'},b^{'})$. Here $E(a,b)$ are the usual expectations of differences in correlation/anti-correlation of the outcomes, where a, a' (b, b') is the
polarizer setting for photon A1' (B1'). In our experiment, we set $a=0$,  $a^{'}=45$, $b=22.5$
and $b^{'}=67.5$, which maximizes the prediction of quantum
mechanics of S to $S_{qm}=2.8$ and leads to a contradiction
between locality and the predictions of quantum mechanics. In our
experiment, the four correlation coefficients between photon 1 and
3 gave the following results: $E(0,22.5)=0.57\pm0.05$, $E(0.67.5)=
-0.67\pm0.04$, $E(45, 22.5)= 0.65\pm0.04$, and $E( 45, 67.5)=
0.69\pm0.04$. Hence $S=2.58\pm0.07$ which violates the classical
limit of 2 by 6 standard deviations. This clearly confirms the
quantum entanglement between the two photons.

In summary, we have presented a feasible and efficient scheme to create
``event ready" entangled photon pairs with single photon sources and detectors. We also provide a proof-in-principle
experimental demonstration of the scheme with the help of
filtered signal photons from down conversion. The generated
``event ready" entangled photon pairs present a strict violation
of Bell's inequality by 6 standard deviations. Although our
experiment is only a proof-in-principle demonstration which still
needs post-selection, the techniques developed in the experiment
can be readily used to generate heralded entangled photon pairs
with the help of photon number detectors \cite{yamamoto3,Fuji},
which will find more application in long distance quantum
communication \cite{brigel} and large scale quantum computation
\cite{KLM,rudolph}.

\begin{acknowledgments}
This work was supported by the NNSF of China, the CAS, the PCSIRT
and the National Fundamental Research Program. This work was also
supported by the Marie Curie Excellent Grant of the EU, the
Alexander von Humboldt Foundation, and the Engineering and Physical Sciences Research Council of the UK.
\end{acknowledgments}


\begin{thebibliography}{99}

\bibitem{conference} Single-Photon Workshop (SPW) 2005: Sources, Detectors,
Applications and Measurement Methods 24 - 26 October 2005,
National Physical Laboratories, UK http://www.spw2005.npl.co.uk/.

\bibitem{njp} New Journal of Physics,
http://www.iop.org/EJ/ abstract/1367-2630/6/1/E04.

\bibitem{qdot} P. Michler \emph{et al}., Science \textbf{290},
2282 (2000); C. Santori, M. Pelton, G. Solomon, Y. Dale, and Y.
Yamamoto, Phys.\ Rev.\ Lett. \textbf{86}, 1502 (2001); V. Zwiller
\emph{et al}., Appl.\ Phys.\ Lett. \textbf{78}, 2476 (2001); Z.
Yuan \emph{et al}., Science \textbf{295}, 102 (2002); J. Vuckovic,
D. Fattal, C. Santor, G. S. Solomon and Y. Yamamoto, Appl.\ Phys.\
Lett. \textbf{82}, 3596 (2003).

\bibitem{nvcentre} C. Kurtsiefer, S. Mayer, P. Zarada and H.
Weinfurter, Phys.\ Rev.\ Lett. \textbf{85}, 290 (2000); A.
Beveratos \emph{et al}., Eur.\ Phys.\ J.\ D \textbf{18}, 191
(2002).

\bibitem{atom} A. Kuhn, M. Hennrich and G. Rempe, Phys.\ Rev.\
Lett. \textbf{89}, 067901 (2002); J. McKeever \emph{et al}.,
Science \textbf{303}, 1992 (2004).

\bibitem{fransonetc} T. B. Pittman, B. C. Jacobs and J. D.
Franson, Phys.\ Rev.\ A \textbf{66}, 042303 (2002).

\bibitem{saw} C. L. Foden, V. I. TAlyanskii, G. J. Milburn, M. L.
Leadbeater and M. Pepper, Phys.\ Rev.\ A \textbf{62}, 011803
(2000).

\bibitem{KLM} E. Knill, R. Laflamme and G. J. Milburn, Nature
\textbf{409}, 46 (2001).

\bibitem{rudolph} D. E. Browne and T. Rudolph, Phys.\ Rev.\ Lett. \textbf{95}, 010501 (2005).

\bibitem{grangier} J. Beugnon \emph{et al}., Nature \textbf{440},
779 (2006).

\bibitem{swapping} M. Zukowski, A. Zeilinger, M. A. Horne and A.
Ekert, Phys.\ Rev.\ Lett. \textbf{71}, 4287 (1993).

\bibitem{jianwei} J.-W. Pan, D. Bouwmeester, H. Weinfurter and A.
Zeilinger, Phys.\ Rev.\ Lett. \textbf{80}, 3891 (1998).

\bibitem{probability} As stated here the probability of the desired condition is only $2/16$. However, the item
$(|HV\rangle_{A1'}|0\rangle_{A2'}|0\rangle_{B1'}|HV\rangle_{B2'}+|0\rangle_{A1'}|HV\rangle_{A2'}|HV\rangle_{B1'}|0\rangle_{B2'})/4$
may collapse photon A1' and B1' into the maximally entangled state
$(|HV\rangle_{A1'}|0\rangle_{B1'}+|0\rangle_{A1'}|HV\rangle_{B1'})/\sqrt{2}$
when there is a coincidence between D1h(D2h) with D1v(D2v), with a
probability of 1/16. We only demonstrate the
former 2/16 probability of success in the experiment.

\bibitem{kwiat} P. G. Kwiat \emph{et al}.,  Phys.\ Rev.\ Lett. \textbf{75}, 4337
(1995).

\bibitem{ou} C. K. Hong, Z. Y. Ou and L. Mandel, Phys.\ Rev.\ Lett. \textbf{59}, 2044 (2005).

\bibitem{Bell} J. S. Bell, Physics (Long Island City, NY) 1, 195
(1964).

\bibitem{chsh} J. F. Clauser, M. Horne, A. Shimony, and R. A.
Holt, Phys.\ Rev.\ Lett. \textbf{23}, 880 (1969).

\bibitem{yamamoto3} E. Waks, K. Inoue, W. D. Oliver, E. Diamanti
and Y. Yamamoto, \emph{IEEE Journal of Selected Topics in Quantum
Electronics} \textbf{9}, 1502(2003).

\bibitem{Fuji} M. Fujiwara and M. Sasaki, Appl.\ Phys.\ Lett.
\textbf{86}, 111119(2005).

\bibitem{brigel} H. -J. Briegel, W. D\"{u}r, J. I. Cirac and P.
Zoller, Phys.\ Rev.\ Lett. \textbf{81}, 5932 (1998).



\end{thebibliography}
\end{document}